\documentclass{elsart}

\usepackage{natbib}
\usepackage{graphicx}

\usepackage{amssymb}

\begin{document}

\begin{frontmatter}



\title{Current Reversals in a inhomogeneous system with asymmetric unbiased fluctuations}


\author[a]{Bao-Quan  Ai}, \author[a]{Xian-Ju Wang},\author[a]{Guo-Tao  Liu},
\author[b]{Hui-Zhang Xie}, \author[a,b]{De-Hua Wen},\author[c]{Wei Chen} and \author[a]{Liang-Gang Liu}
\address[a]{Department of Physics, ZhongShan University,
GuangZhou, China}
\address[b]{Department of Physics, South China University of
technology, GuangZhou, China}
\address[c]{Department of Physics, JiNan  University, GuangZhou,
China}

\begin{abstract}
\indent We present a study of transport of a Brownian particle
moving in periodic symmetric potential in the presence of
asymmetric unbiased fluctuations. The particle is considered to
move in a medium with periodic space dependent friction. By tuning
the parameters of the system, the direction of current exhibit
reversals, both as a function of temperature as well as the
amplitude of rocking force. We found that the mutual interplay
between the opposite driving factors is the necessary term for
current reversals.

\end{abstract}


\end{frontmatter}

\section{Introduction}
\label{1}




\indent Recently, there has been increasing interest in studying
the noise-induced transport of Brownian particles for systems with
a spatially periodic potential field. It has been shown that
asymmetry of the potential \cite{1}\cite{2}, the asymmetry of the
driving noise \cite{3}, and the input signal\cite{4} are
ingredients for the transport. These subjects were motivated by
the challenge to explain undirection of transport in biological
systems, and several models have been proposed to describe
muscle's contraction\cite{5}\cite{6}\cite{7}, or the asymmetric
polymerization of actin filaments responsible of cell
motility\cite{1}.

\indent Rectification of noise leading to unidirectional motion in
ratchet systems has been an active field of research over the last
decade. In these systems directed Brownian motion of particles is
induced by nonequilibrium noise in the absence of any net
macroscopic forces and potential gradients. Several physical
models have been proposed: rocking ratchets \cite{8}, fashing
ratchets \cite{9}, diffusion ratchets \cite{10}, correlation
ratchets \cite{11}, etc. In all these studies the potential is
taken to be asymmetric in space. It has also been shown that one
can obtain unidirectional current in the presence of spatially
asymmetric potentials. For these nonequilibrium systems external
random force should be time asymmetric or the presence of space
dependent mobility is required.

\indent The study of current reversal phenomena has given rise to
research activity on its own. The motivation being possibility of
new particle separation devices superior to existing methods such
as electrophoretic method for particles of micrometer
scale\cite{12}. It is known that current reversals in ratchet
systems can be engendered by changing various system
parameters\cite{13}\cite{14}, including flatness parameter of the
noise\cite{15}, the correlation time of nonequilibrium
fluctuations\cite{16}, the temperature in multinoise
cases\cite{17}, the power spectrum of the noise source\cite{18},
the shape of the potential\cite{19}, the number of interacting
particles per unit cell\cite{20} and the mass of the
particles\cite{21}. In this paper, we study the current of a
Brownian particle in periodic symmetric potential in the presence
of asymmetric unbiased fluctuation and the inhomogeneous friction
and show when the current reversals occur.

\section{The current in an inhomogeneous system}
\indent The Brownian  dynamics of the overdamped  particle moving
under the influence of a symmetric potential $V_{0}(x)$ and
subject to a space dependent friction coefficient $\gamma(x)$ and
asymmetric unbiased fluctuations at temperature $T$, is described
by the Langevin equation\cite{11}
\begin{equation}\label{1}
    \frac{dx}{dt}=-\frac{V^{'}_{0}(x)-F(t)}{\gamma(x)}-k_{B}T\frac{\gamma^{'}(x)}{[\gamma(x)]^{2}}+\sqrt{\frac{k_{B}T}{\gamma(x)}}\xi(t),
\end{equation}
where $\xi(t)$ is randomly fluctuating Gaussian white noise with
zero mean and correlation: $<\xi(t)\xi(t^{'})>=2\delta(t-t^{'})$.
Here $<...>$ denotes an ensemble average over the distribution of
the fluctuating forces $\xi(t)$. The primes in the Eq. (1) denote
the derivative with respect to the space variable $x$. It should
be noted that the Eq. (1) involves a multiplicative noise with an
additional temperature dependent drift term. This term turns out
to be important in order for the system to approach the correct
thermal equilibrium sate. We take
$V_{0}(x)=V_{0}(x+2n\pi)=-\sin(x)$, $n$ being any natural integer.
Also, we take the friction coefficient $\gamma(x)$ to be periodic:
$\gamma(x)=\gamma_{0}(1-\lambda\sin(x+\phi))$, Where $\phi$ is the
phase different with respect to $V_{0}(x)$. The evolution of the
probability density for $x$ is given by Fokker-Planck
equation\cite{1}
\begin{equation}\label{2}
    \frac{\partial P(x,t)}{\partial t}=\frac{\partial}{\partial
    x}\frac{1}{\gamma(x)}[k_{B}T\frac{\partial P(x,t)}{\partial
    x}+(V^{'}_{0}(x)-F(t))P(x,t)]=-\frac{\partial j}{\partial x} .
\end{equation}
where $j$ is the probability current and it can be expressed as
follows:
\begin{equation}\label{3}
    j(x,t)=-\frac{1}{\gamma(x)}[(V^{'}_{0}(x)-F(t))+k_{B}T\frac{\partial}{\partial x}]P(x,t).
\end{equation}
If $F(t)$ changes very slowly, there exists a quasi-stationary
state. In this case, the average current  of the particle can be
solved by evaluating the constants of integration under the
normalization condition and the periodicity condition of $P(x)$,
and the current can be obtained and expressed as
\begin{equation}\label{4}
j(t)=\frac{k_{B}T(1-e^{2\pi
F(t)/k_{B}T})}{\int^{2\pi}_{0}dy\exp(-F(t)y/k_{B}T)C(y)},
\end{equation}
where the space correlation function is given by
\begin{equation}\label{5}
C(y)=\frac{1}{2\pi}\int^{2\pi}_{0}dx\gamma(x+y)\exp(-\frac{V_{0}(x)-V_{0}(x+y)}{k_{B}T}).
\end{equation}
\indent Considering that the external force $F(t)$ is slowly
changing with the time, the average probability current $J$ over
the time interval of a period can expression by
\begin{equation}\label{6}
J=\frac{1}{\tau}\int^{\tau}_{0}j(F(t))dt.
\end{equation}
where $\tau$ is the period of the driving force $F(t)$, which is
assumed longer than any other time scale of the system in this
adiabatic limit. Here we will consider a driving with a zero mean
$<F(t)>=0$, but which is asymmetric in time\cite{22}, as shown in
Fig.1.
\begin{eqnarray}\label{7}
  F(t)=\frac{1+\varepsilon}{1-\varepsilon}F,               (n\tau\leq
t<n\tau+\frac{1}{2}\tau(1-\varepsilon)), \\
      =-F,    (n\tau+\frac{1}{2}\tau(1-\varepsilon)<t\leq
      (n+1)\tau).
\end{eqnarray}

\begin{figure}[htbp]
  \begin{center}\includegraphics[width=10cm,height=6cm]{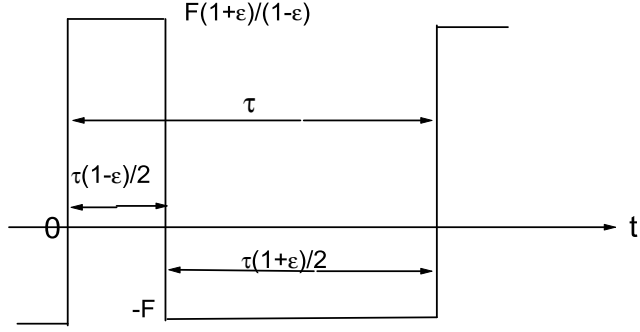}
  \caption{The driving force $F(t)$ which preserved the zero
mean$<F(t)>=0$, where the temporal asymmetry is given by the
parameter $\varepsilon$.}\label{1}
\end{center}
\end{figure}

 \indent In this case the time average current is easily
calculated,
\begin{equation}\label{9}
    J=\frac{1}{2}[(1-\varepsilon)j(\frac{(1+\varepsilon)F}{1-\varepsilon})+(1+\varepsilon)j(-F)].
\end{equation}
\section{Results and Discussion}
\indent We have calculated the average current about the motion of
the Brownian particle in a periodic symmetric potential with the
asymmetric unbiased fluctuations.

\begin{figure}[htbp]
  \begin{center}\includegraphics[width=11cm,height=8cm]{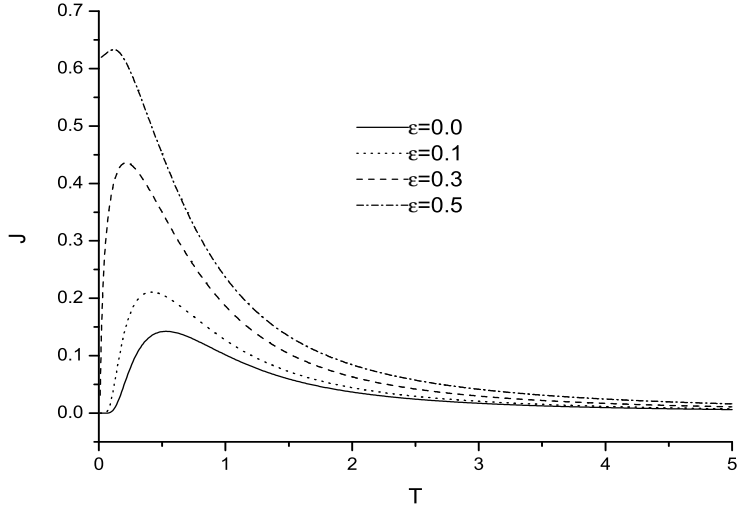}
  \caption{The net current $J$ versus temperature $T$ for different
values of $\varepsilon$ ($\varepsilon>0$). $\lambda$=0.9,
$\gamma_{0}$=0.1, $F$=0.5 and $\phi$=1.3$\pi$.}\label{2}
\end{center}
\end{figure}

\begin{figure}[htbp]
  \begin{center}\includegraphics[width=11cm,height=8cm]{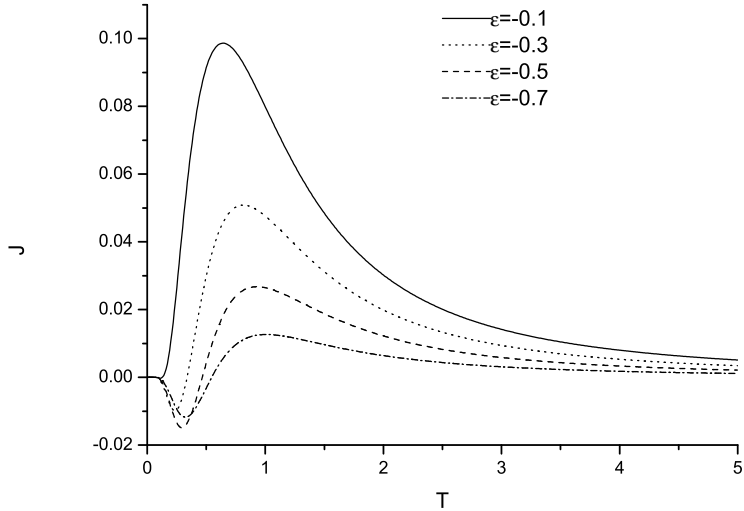}
  \caption{The net current $J$ versus temperature $T$ for different
values of $\varepsilon$ ($\varepsilon<0$) at  The other parameters
is the same as the Fig. 2.}\label{3}
\end{center}
\end{figure}
\indent The average current $J$ is plotted in Fig. 2 as a function
of temperature $T$ for different asymmetry parameters
($\varepsilon$ is positive ). Here $\lambda$=0.9,
$\gamma_{0}$=0.1, $F$=0.5 and $\phi$=1.3$\pi$. The figure shows
that the average current is a peaked function of the temperature.
With increasing of intensity of the asymmetry parameter the
maximal current increases, but the corresponding temperature at
which the current takes the maximum is shifted to the lower
temperature. For very high temperature the current vanishes just
like that of most o the thermal ratchet models
\cite{1}\cite{2}\cite{3}. It is obvious that no current reversals
occur at the case $\phi>\pi$ and $\varepsilon>0$. In Fig. 3 we
plotted the average current vs temperature for different asymmetry
parameters ($\varepsilon<0$). The other parameters is the same as
the Fig.2. From Fig. 3, we can see that when the asymmetry
parameter is negative the current reversal may occur
($\varepsilon$=-0.3,  -0.5, -0.7) and the current vanishes for
lower temperature as well as higher temperature.

\begin{figure}[htbp]
  \begin{center}\includegraphics[width=11cm,height=8cm]{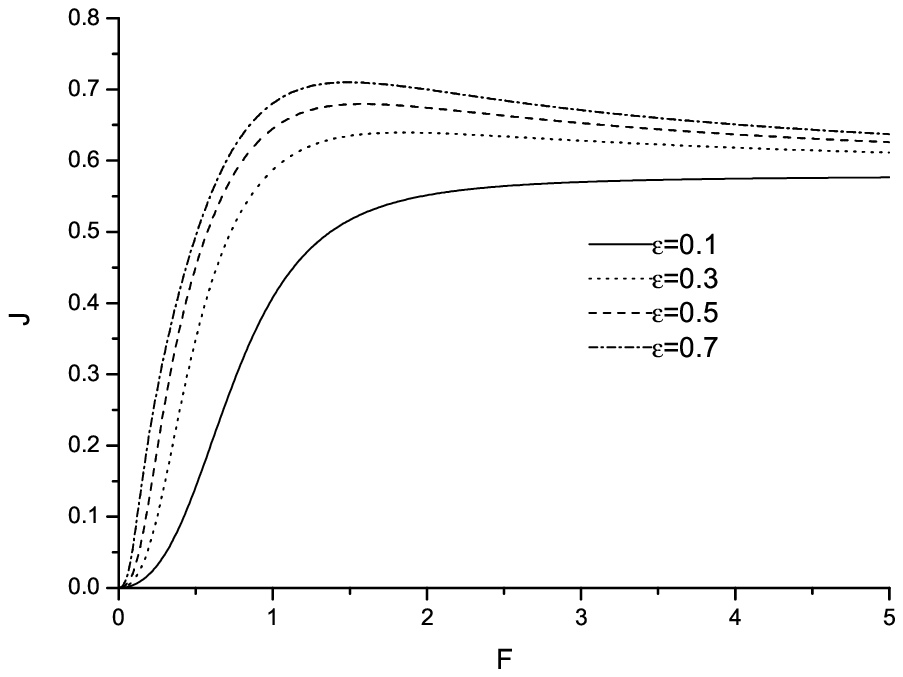}
  \caption{The net current $J$ versus rocking force $F$ for different
values of $\varepsilon$ ($\varepsilon>0$). $\lambda$=0.9,
$\gamma_{0}$=0.1, $T$=0.5 and $\phi$=1.3$\pi$.}\label{4}
\end{center}
\end{figure}

\begin{figure}[htbp]
  \begin{center}\includegraphics[width=11cm,height=8cm]{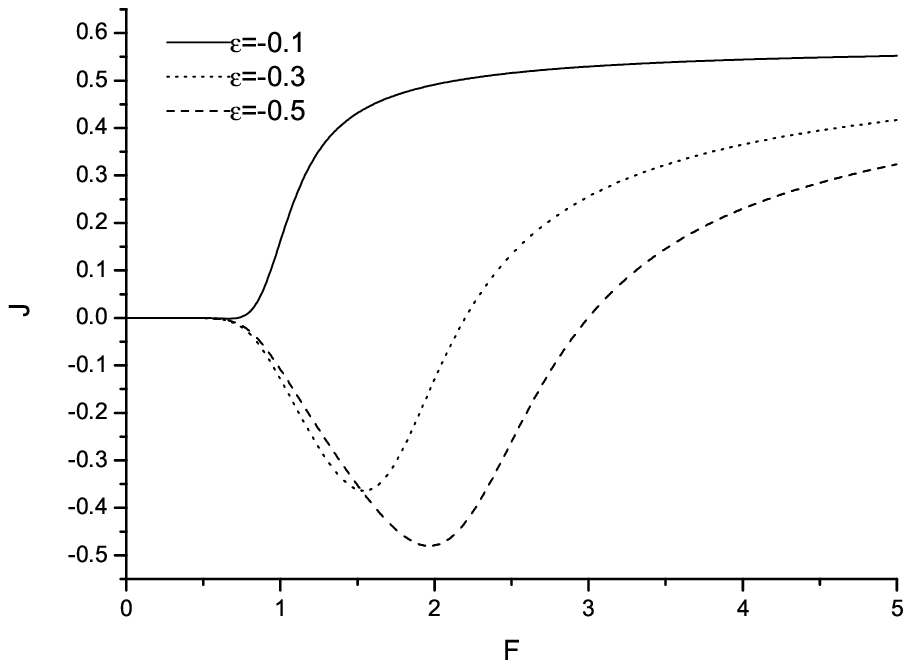}
  \caption{The net current $J$ versus rocking force $F$ for different
values of $\varepsilon$ ($\varepsilon<0$). The other parameters is
the same as the Fig. 4.}\label{5}
\end{center}
\end{figure}
\indent In Fig. 4, the plot of $J$ versus $F$ is shown for
different asymmetry parameters ($\varepsilon$ is positive),
keeping $\lambda$, $\gamma_{0}$, $T$ and $\phi$ fixed 0.9, 0.1,
0.5 and 1.3$\pi$ respectively. With increasing of the asymmetry
parameters, the current increases. For very large value of $F$,
the current asymptotically goes to a positive constant value
depending on the value $\phi$, as was previously shown for the
adiabatic case \cite{14}. In the absence of space dependent
friction ratchets where the currents saturate to zero in the same
asymptotic regime. It is obvious that there are no current
reversals in the case $\phi>\pi$, $\varepsilon>0$. But the current
reversals vs $F$ may occur (see Fig. 5) when the asymmetry
parameters are negative ($\varepsilon=-0.3$, $\varepsilon=-0.5$).
In the present case (Fig. 5) $\varepsilon$ is chosen so that the
current goes to negative direction at lower temperature and
positive direction at higher temperature which guarantees the
current reversal occurs.

\begin{figure}[htbp]
  \begin{center}\includegraphics[width=11cm,height=8cm]{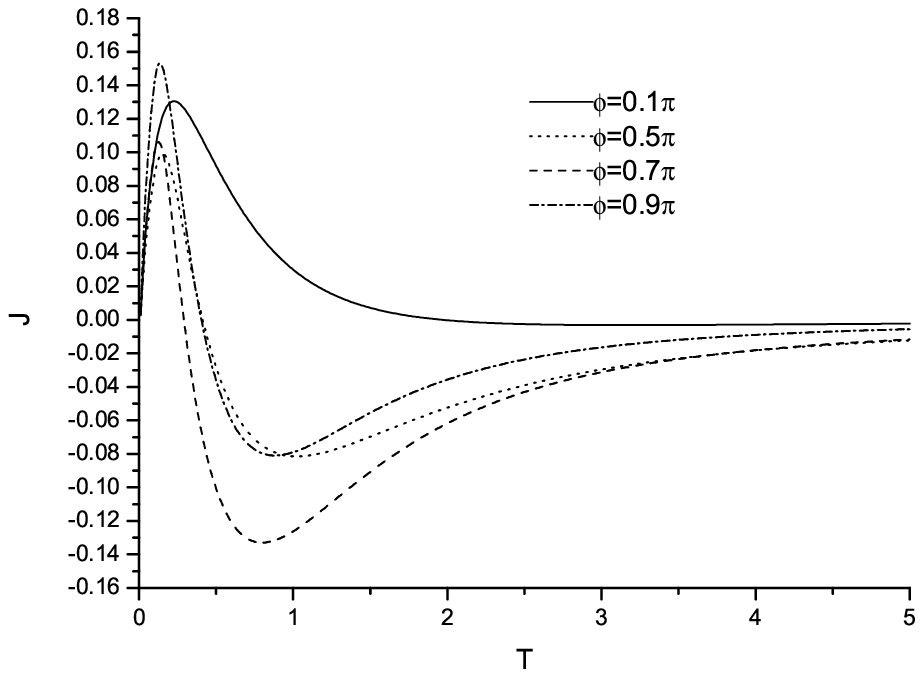}
  \caption{The net current $J$ versus temperature $T$ for different
values of $\phi$. $\lambda$=0.9, $\gamma_{0}$=0.1, $F$=0.5 and
$\varepsilon$=0.3.}\label{6}
\end{center}
\end{figure}

\begin{figure}[htbp]
  \begin{center}\includegraphics[width=11cm,height=8cm]{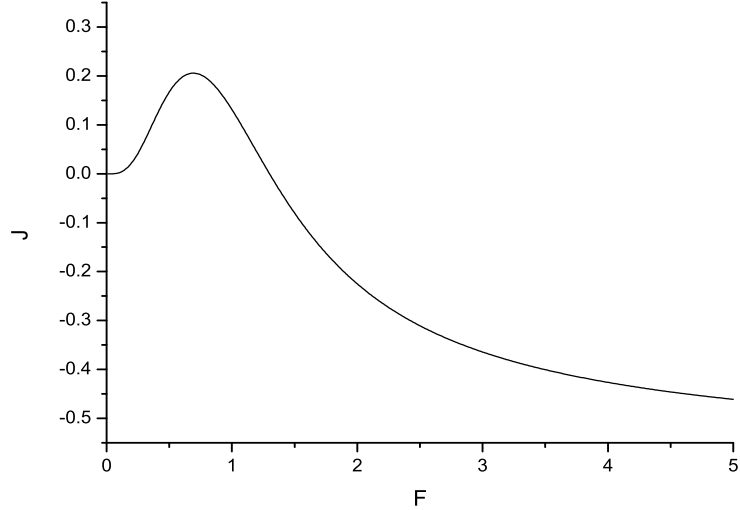}
  \caption{The net current $J$ versus rocking force $F$ for
$\phi=0.5\pi$, $\lambda$=0.9, $\gamma_{0}$=0.1, $T$=0.5 and
$\varepsilon$=0.3.}\label{7}
\end{center}
\end{figure}
\indent In Fig. 6  we plotted $J$ for various values of $\phi$
($\phi<\pi$) at $\varepsilon$=0.3 as a function of temperature
$T$. Here $\lambda$=0.9, $\gamma_{0}$=0.1 and $F$=0.5. When
$0<\phi<\pi$ and $\varepsilon$=0.3($\varepsilon>0$), the current
reversal may occur and the current saturates to zero for high
temperature, namely in high temperature the ratchet effect
disappears. In the Fig. 7  the average current $J$ is plotted vs
$F$ for $\phi$=0.5$\pi$ and $\varepsilon$=0.3. The other
parameters is the same as the Fig. 6.  From the Fig. 7 we can see
that the current reversal vs $F$ occurs at the case
$\phi$<$\pi$,$\varepsilon$>0. The current saturates to a negative
constant for large value of $F$ which is similar to the Fig. 4

\indent When the fluctuation is temporally asymmetric, its
correlation properties in either direction are different, and a
net current can arise even in the absence of a spatial asymmetry
and a space dependent friction\cite{22}. The positive asymmetric
parameters induce a positive direction current while the negative
parameters give a negative directional current. On the other hand,
the phase difference between the friction coefficient and the
symmetric potential is sensitive to direction of the net current.
Even in a symmetric potential and symmetric fluctuations\cite{14}
a net current can arise. The current tends to positive direction
for $\varphi>\pi$ and negative direction for $0<\phi<\pi$. In
fact, our ratchet contains these two driving factor. It is found
that the current reversals may occur when a negative driving
factor meets a positive driving factor: Case A ($\varepsilon<0$
and $\phi>\pi$ see Fig.3, Fig.5), Case B ($\varepsilon>0$ and
$0<\phi<\pi$ see Fig.6, Fig.7), while no current reversals occur
when the two negative (positive) driving factors meet: Case C
($\varepsilon>0$ and $\phi>\pi$ see Fig. 2, Fig. 4), Case D
($\varepsilon<0$ and $0<\phi<\pi$).

  \indent In a word, in our symmetric potential ratchet,
  $\varepsilon\phi<0$ is the necessary condition for current
  reversals, the particle never changes its moving direction under
  the condition of $\varepsilon\phi>0$ and even no current occur
  under the condition of $\varepsilon\phi=0$.

  \section{Summary and Conclusion}
\indent In present paper, the transport of a Brownian particle
moving in spatially symmetric potential in the presence of an
asymmetric unbiased fluctuation is investigated.  The current of
the ratchet is discussed for different cases.  We find that the
mutual interplay between the opposite driving factors is the
necessary term for current reversals. We find current reversal,
both as a function of temperature as well as the amplitude of
rocking force, when the force is adiabatic and the potential is
symmetry.

\indent The phenomena of current reversals may be interest in
biology, e. g., when considering the motion of macromolecules. It
is known that the two current reversals effect allows one pair of
motor proteins to move simultaneously in opposite directions along
the microtubule inside the eucariotic cells.

\indent To summarize, it is remarkable that the interplay of
asymmetric unbiased fluctuation, in homogeneous friction and
thermal noise with spatially symmetric potential generates such a
rich variety of cooperation effects as up to current reversals
with temperature as well as the rocking forcing.

{\bf Acknowledgements}\\
 \indent The project supported by National
Natural Science Foundation of China (Grant No. of 10275099) and
GuangDong Provincial Natural Science Foundation (Grant No. of
021707 and 001182).\\









\end{document}